\documentclass[aps,prstab,twocolumn,groupedaddress,nofootinbib,floatfix]
{revtex4-1}

\usepackage{amsmath}
\usepackage{graphicx}
\usepackage[caption=false]{subfig}
\usepackage{color,soul}

\usepackage{relsize}
\usepackage{xspace}

\newcommand{\unit}[1]{\ensuremath{\,\mathrm{#1}}}

\begin{document}


\title{
Practical aspects of transverse resonance island buckets at the Cornell Electron Storage Ring: design, control and application
}



\author{S. T. Wang}
\email[]{sw565@cornell.edu}
\author{V. Khachatryan}
\affiliation{Cornell Laboratory for Accelerator-based Sciences and 
  Education, Cornell University, Ithaca, NY 14853}




\begin{abstract}
In an accelerator, the nonlinear behavior near a horizontal resonance line ($n\nu_x$) usually involves the appearance of stable fixed points (SFPs) in the horizontal phase space, also referred to as transverse resonance island ``buckets" (TRIBs). Specific conditions are required for TRIBs formation. At the Cornell Electron Storage Ring, a new method is developed to improve the dynamic and momentum apertures in a 6-GeV lattice as well as to preserve the conditions for TRIBs formation. This method reduces the dimension of variables from 76 sextupoles to 8 group variables and then utilizes the robust conjugate direction search algorithm in optimization. Created with a few harmonic sextupoles or octupoles, several knobs that can either rotate the TRIBs in phase space or adjust the actions of SFPs are discussed and demonstrated by both tracking simulations and experimental results. In addition, a new scheme to drive all particles into one single island is described. Possible applications using TRIBs in accelerators are also discussed.
\end{abstract}


\maketitle

\section{Introduction}\label{intro}
In recent years, the nonlinear resonances forming stable islands in phase space, also referred to as transverse resonance island ``buckets" (TRIBs), gain more interests in the light source and x-ray science community due to its distinct way of separating the x-ray pulses in both time and space \cite{bessy:2015, bessy:2019, max4:2019, hel_switch:2020, max4:2021, spear3:2022, chess_tribs:ipac2022, sci_rep:2022,chess_tribs:ipac2023, chess_tribs:2023, pls:2024, chess_tribs:ipac2024}. For example, at BESSY-II, the stable two-orbit operation utilizing the TRIBs to conduct unique x-ray experiments have been demonstrated \cite{bessy:2019, hel_switch:2020, sci_rep:2022}. Progresses have also been made on the theoretical understanding of formation of TRIBs, advancing the tools to manipulate the TRIBs in phase space for better control the x-ray pulses from TRIBs \cite{chess_tribs:2023}. However, there has been a long history of utilizing the resonance nature of synchrotrons and storage rings, such as to extract particles in multi-turns at the CERN-PS and many other facilities \cite{cern:pac1993, cern:2002, cern:2016}. This resonant extraction technique has been widely implemented to extract hardons for high energy fixed-target experiments or ion therapy \cite{therapy:1996}. A recent effort is also made to study the extraction of electrons from low emittance booster rings using TRIBs at DESY \cite{desy:2024}. Thus it is evident that understanding TRIBs and developing techniques to control this unique nonlinear behavior are still necessary and useful for current and future applications.

At the Cornell Electron Storage Ring (CESR), Hamiltonian perturbation theory and map-based PTC method have been successfully implemented to design a TRIBs lattice and predict the fixed points near a third-order ($3\nu_x$) resonance \cite{chess_tribs:2023}. Note PTC refers to ``fully polymorphic package"-``polymorphic tracking code" library which handles Taylor map manipulation and Lie algebraic operations \cite{ef_book:1997, ef_book:2016}. Subsequent experiments confirmed the observed TRIBs locations near $3\nu_x$ as well as the $4\nu_x$ resonance, which agree reasonably well with theoretical calculation and simulations \cite{chess_tribs:2023, chess_tribs:ipac2024}. While exploring the TRIBs experimentally in CESR, we have faced a few challenges such as experiencing poor lifetime while adjusting the horizontal tune near the resonance or with application of a sinusoidal kick, indicating limited dynamic aperture (DA) or momentum aperture (MA) near the resonance. Since the TRIBs lattice intrinsically contains large nonlinearity, conventional methods involving reducing the nonlinearity of the lattice in order to improve DA do not work for the TRIBs lattice. Besides improving beam lifetime and injection with improved DA, other practical subjects such as how to systematically tune the angles and locations of islands and driving particles into one single island have also been explored at CESR.

In this paper, we will discuss a new method to improve DA and MA in the TRIBs lattice while preserving the properties of TRIBs. This method reduces the dimension of variables from 76 sextupoles to 8 DA knobs and then utilizes the robust conjugate direction search (RCDS) algorithm \cite{rcds:2013} to optimize the DA or MA directly. Simulation shows the MA is improved significantly after optimization, confirmed by the experiments with improved lifetime. We will also discuss three knobs consisting of a few harmonic sextupoles or octupoles, which can either rotate the TRIBs in phase space or adjust the actions of stable fixed points (SFPs). Their effects are demonstrated by both tracking simulations and experiments. These practical knobs will be useful for fine tuning of x-ray pulses. Besides manipulating the TRIBs in phase space, a new scheme to drive all particles into one single island is described. Several potential applications using TRIBs in accelerators are also discussed.

The paper is organized as follows. CESR lattice is briefly introduced in Sec.~\ref{lat}. The method to improve the DA and MA is discussed in Sec.~\ref{ma_opt} while phase-space-control knobs are introduced and discussed in Sec.~\ref{phase_con}. In Sec.~\ref{single}, we then describe the new driving scheme that pushes all particles into one single island and explain the reason behind it. Finally, we discuss current and two potential applications in accelerators using TRIBs and conclude in Sec.~\ref{app}.

\begin{figure}
   \centering
   \includegraphics*[width=220pt]{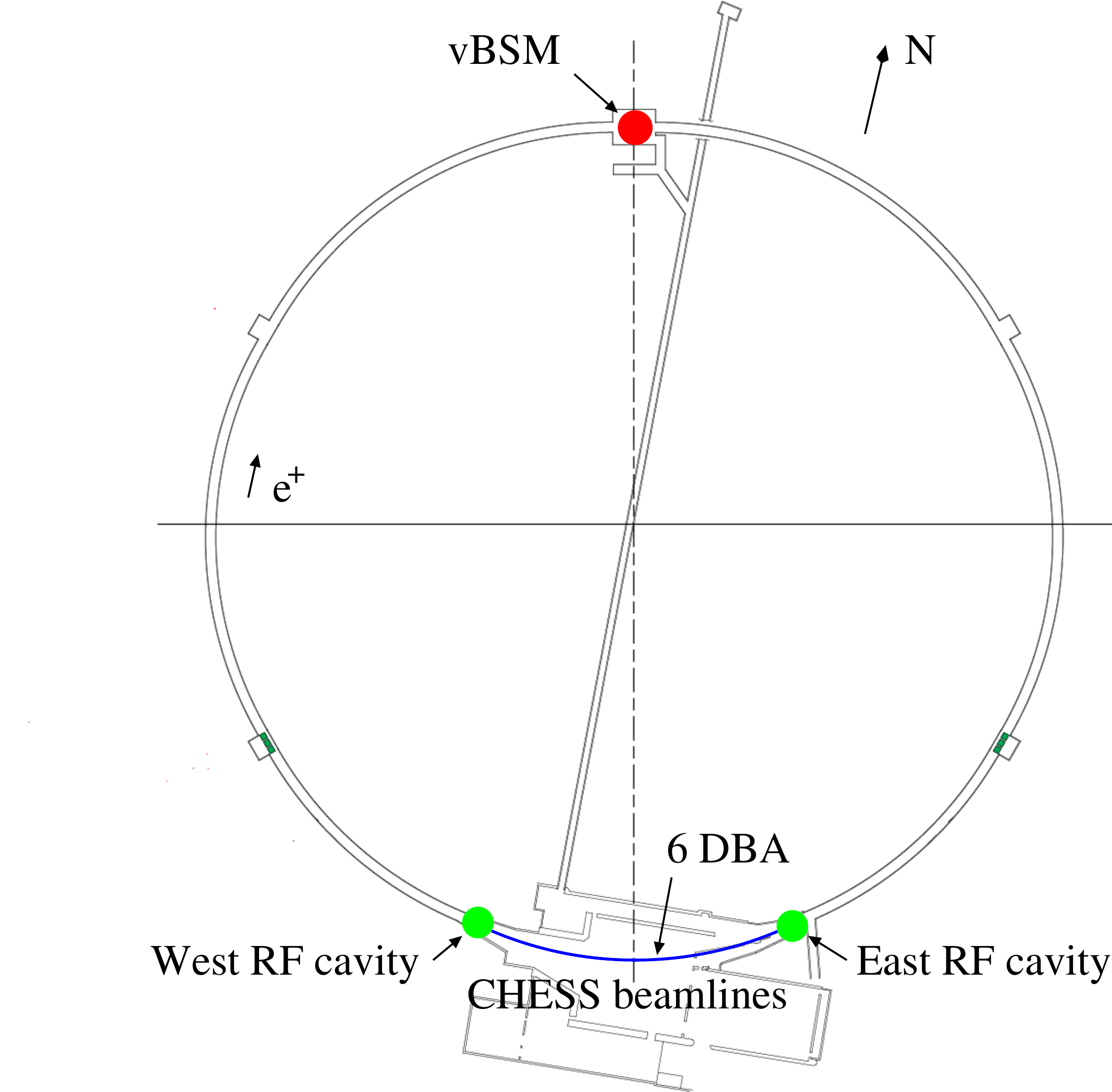}
   \caption{CESR layout showing the locations of RF cavities, CHESS beamlines, and visible-light beam size monitor (vBSM).}
   \label{fig:layout}
\end{figure}

\section{CESR lattices}\label{lat}
CESR, built on the Cornell University campus, stores counterrotating beams of electrons and positrons and has been operated as a collider for high-energy physics experiments for many decades. Since 2008, CESR has served as a dedicated light source for x-ray users, namely Cornell High Energy Synchrotron Source (CHESS). The x-ray beamlines are located in the southern area of the ring as shown in Fig.~\ref{fig:layout}. In 2018, one sextant of the ring between the east and west RF cavities was upgraded with double bend acromat (DBA) to reduce the ring emittance and accommodate more compact undulators while the rest of the ring remains as FODO lattice \cite{chessu:2019}. After the CHESS upgrade (CHESS-U), the main accelerator parameters for CHESS operation are listed in Table~\ref{table1}. CESR magnets including 113 quadrupoles, 12 dipole quadrupoles (combined function dipoles), and 76 sextupoles are all individually powered, which offers significant flexibility for lattice design and complex nonlinear dynamics studies. Only five harmonic sextupoles exist while two of them are near the east RF cavity and three are near the west cavity. The visible-light beam size monitor (vBSM), which is extensively used to view and acquire images of islands during TRIBs experiments, is located in the north area of CESR \cite{vbsm:2013}.  Unlike most 3rd and 4th generation light sources, there is no global periodicity in the CESR lattice. Details of CHESS-U lattice can be found in Ref. \cite{chessu:2019}.

\begin{table}[t]
   \centering
   \caption{CESR Parameters for CHESS-U operation}
   \begin{tabular}{lcr}
   \hline
Beam energy (GeV)             &$E_0$              & $6.0$\\
Circumference (m)             &$L$                &$768.438$\\
Transverse damping time (ms)  &$\tau_{x,y}$             &$12.0$, $14.6$\\
\quad\quad\quad\quad\quad\quad\quad\quad\quad\quad\quad(turns)  &                &4685, 5700\\
Longitudinal damping time (ms)  &$\tau_z$             & $8.2$\\
Horizontal tune               &$\nu_x$              & $16.556$\\
Vertical tune                 &$\nu_y$              & $12.636$\\
Synchrotron tune              &$\nu_s$              & $0.027$\\
Horizontal emittance (nm rad)  &$\epsilon_x$     &$\sim28$\\
Energy spread                  &$\sigma_p$        & $8.2\times10^{-4}$\\     
Revolution frequency (kHz)          &$f_{rev}$         & $390.14 $  \\
   \hline
   \end{tabular}
   \label{table1}
\end{table}

\begin{figure}[b]
   \centering
   \includegraphics*[width=240pt]{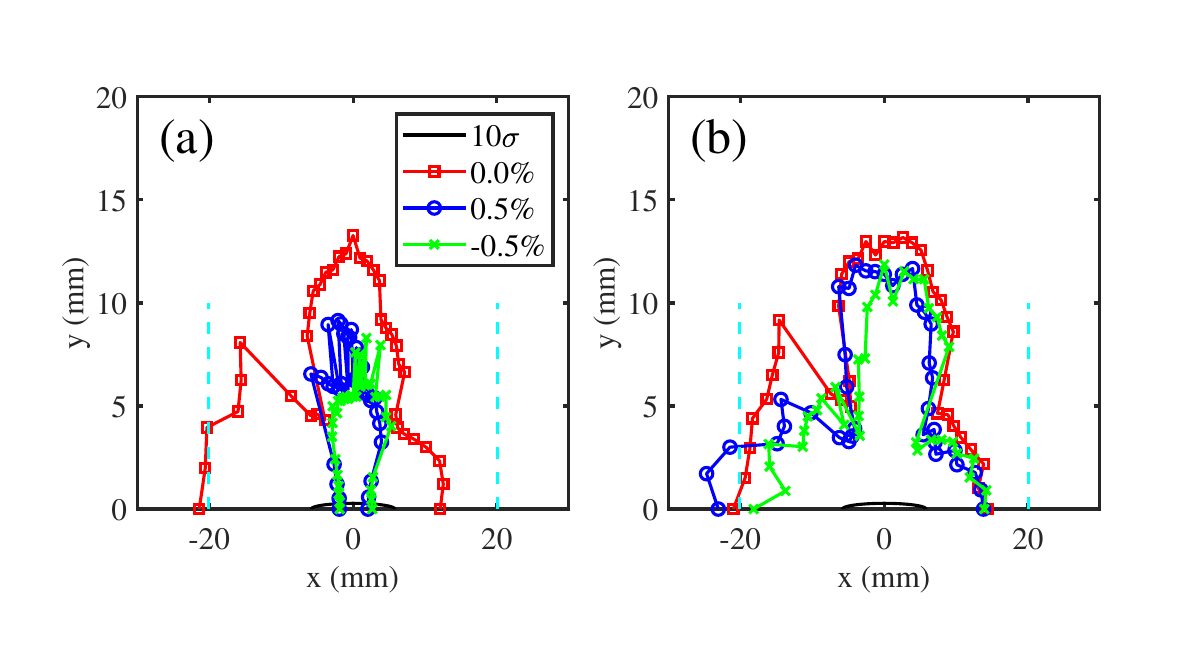}
   \caption{The dynamic apertures before (a) and after (b) the sextupole optimization. The black solid line shows the 10$\sigma$ beam eclipse. The cyan dashed line indicates the projected horizontal physical aperture.}
   \label{fig:da_ma}
\end{figure}

As shown in Table \ref{table1}, the nominal horizontal fractional tune ($Q_x$) during CHESS operation is $0.556$, far from the 3rd-order resonance ($\sim$0.667). In this standard CHESS-U lattice, TRIBs do not appear by only adjusting the tune to near $0.667$. Therefore, in order for TRIBs to appear, a 6-GeV lattice with the horizontal tune near a 3rd-order resonance at $3\nu_{x}=50$ was designed and the sextupoles were carefully optimized to achieve desired resonant driving terms (RDTs), especially $h_{30000}$ and $h_{22000}$ \cite{bengtsson:1997}. Details of designing TRIBs lattice can be found in Ref.~\cite{chess_tribs:2023}. After a few passes of linear and nonlinear lattice optimization, although the on-energy DA is good, comparable to the physical aperture, the off-energy DAs or the MA are poor, even less than the $10\sigma$ beam ellipse as shown in Fig.~\ref{fig:da_ma} (a). The $10\sigma$ beam ellipse is plotted from the equation $\frac{x^2}{(10\sigma_x)^2}+\frac{y^2}{(10\sigma_y)^2}=1$, where $\sigma_x$ and $\sigma_y$ are the horizontal and vertical beam sizes respectively assuming $1\%$ $xy$ coupling. In Fig.~\ref{fig:da_ma}, the DAs are acquired by tracking particles for 2000 turns with initial coordinates ($x_0$, 0, $y_0$, 0, 0, $\sigma_e$), where ($x_0$, $y_0$) are along 37 azimuth angles in the $xy$ plane with a $0.5$-degree step and the energy offset $\sigma_e$ is set at $0$, $0.5\%$ and $-0.5\%$, respectively. The points plotted in Fig.~\ref{fig:da_ma} are the boundaries beyond which the particles are lost within 2000 turns, indicating the DAs. All the simulation programs discussed in this paper are based on the BMAD code library \cite{bmad:2006}.

During initial studies of the TRIBs lattice, the stable islands were observed with $Q_x$ near the 3rd-order resonance however beam lifetime was poor while adjusting $Q_x$ near the resonance especially after application of a sinusoidal kick to drive particles into islands. Apparently, this behavior can be explained by poor DA or MA, which needs improvement.

\section{Momentum aperture optimization}\label{ma_opt}

For standard CHESS-U lattice or other conventional lattices, the methods of optimizing sextupoles to improve DA and MA normally involve reducing the nonlinarity in the lattice, such as optimizing the determinant of the 1-turn Jacobian matrix \cite{dlr:2001}, minimizing RDTs \cite{bengtsson:1997}, square matrix method \cite{yu:2017}, or suppressing the chaos \cite{entropy:2024}. However, these methods would not apply for the TRIBs lattice since the TRIBs lattice intrinsically contains strong nonlinearity and requires relatively large amplitude dependent tune shift (ADTS). Thus, direct DA optimization is chosen for our TRIBs lattice while still preserving the conditions for TRIBs formation.

\begin{figure}
   \centering
   \includegraphics*[width=240pt]{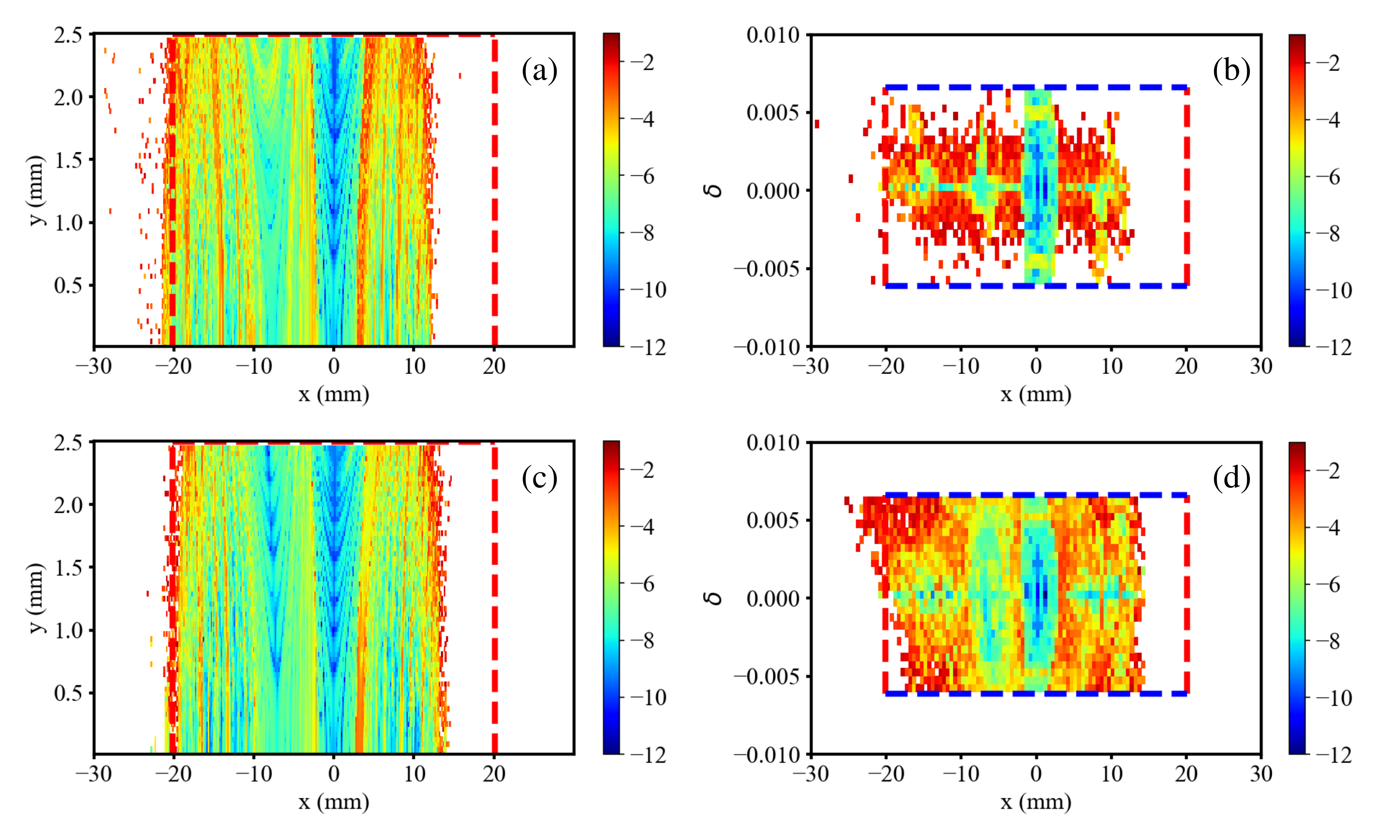}
   \caption{Frequency map analysis results show the DA and MA before ((a) and (b)) and after ((c) and (d)) the sextupole optimization. The red dashed lines indicate the projected physical apertures and the blue dashed lines indicate CESR RF momentum acceptance.}
   \label{fig:fmap}
\end{figure}

In CESR there are a total of 76 sextupoles, which are all individually powered. A large horizontal orbit displacement bump is introduced near the injection point to facilitate injection. In order to avoid undesired large effects from varying the sextupoles inside the horizontal orbit bump, we have excluded the 9 sextupoles in that region, leaving us with 67 sextupoles available for tuning. It may not be efficient and practical to include these 67 variables in the optimization process. Therefore, we apply the dimension reduction method to reduce the variables from 67 sextupoles to 8 knobs. The method has been described in detail in Refs.~\cite{wb_paper:2019,wb_thesis:2020}. Here we outline the critical steps as follows. The first step is to create a direct DA Hessian matrix $H$, the second derivative of the objective (DA) with respect to the 67 sextupoles. In the calculation of Hessian matrix, DA is obtained from tracking and quantified using the sum of the squares of the extent of the on-energy DA in the positive and negative horizontal directions with no vertical offset. While tuning the DA knobs, we also want to avoid changing the horizontal and vertical chromaticities ($h_{11001}$, $h_{00111}$) as well as the 3rd-order resonance ($h_{30000r}$, $h_{30000i}$) and ADTS ($h_{22000r}$, $h_{11110r}$, $h_{00220r}$) terms so as to preserve the TRIBs properties. Note all the RDTs are defined in Ref. \cite{bengtsson:1997}. Hence, the second step is to find the $67\times60$ matrix $Q$ whose columns are the null-space vectors for the 7 RDTs from the Jacobian matrix of them with respect to the 67 sextupoles. Finally, a modified Hessian matrix $\hat{H}$ can be created by $\hat{H}=Q^{T}HQ$, the eigenvectors $\hat{E}$ of $\hat{H}$ are found by singular value decomposition (SVD) analysis, and total 60 knobs ($Q\hat{E}$) will be created. Although in principle all 60 knobs can be used in the optimization, we find the eight with the largest eigenvalues to be adequate for effective optimization.

In terms of the optimization process, there are many algorithms available including Levenberg-Marquardt least-squares (LM), differential evolution (DE), genetic algorithm (GA), etc. We explored LM and DE methods in Tao \cite{tao:2005} to optimize the DA and both worked well as expected. The RCDS method, commonly used for online optimization \cite{rcds:2013}, proved effective for offline optimization of the DA/MA for our TRIBs lattice. Here we only show the results with RCDS method. Since the off-energy DA is the culprit as shown in Fig.~\ref{fig:da_ma} (a), the negative value of the area enclosed by the off-energy DA curve ($\delta_E=-0.5\%$) was chosen as the optimization objective. The first 8 DA knobs described above were considered as the variables in the optimization. After $\sim$400 evaluations, the objective reaches minimum. The best case with the largest DA area (minimum objective) was selected and its DAs is shown in Fig.~\ref{fig:da_ma} (b). As we can see, both off-energy DAs have been greatly improved and the on-energy DA is improved slightly.

\begin{figure}
   \centering
   \includegraphics*[width=240pt]{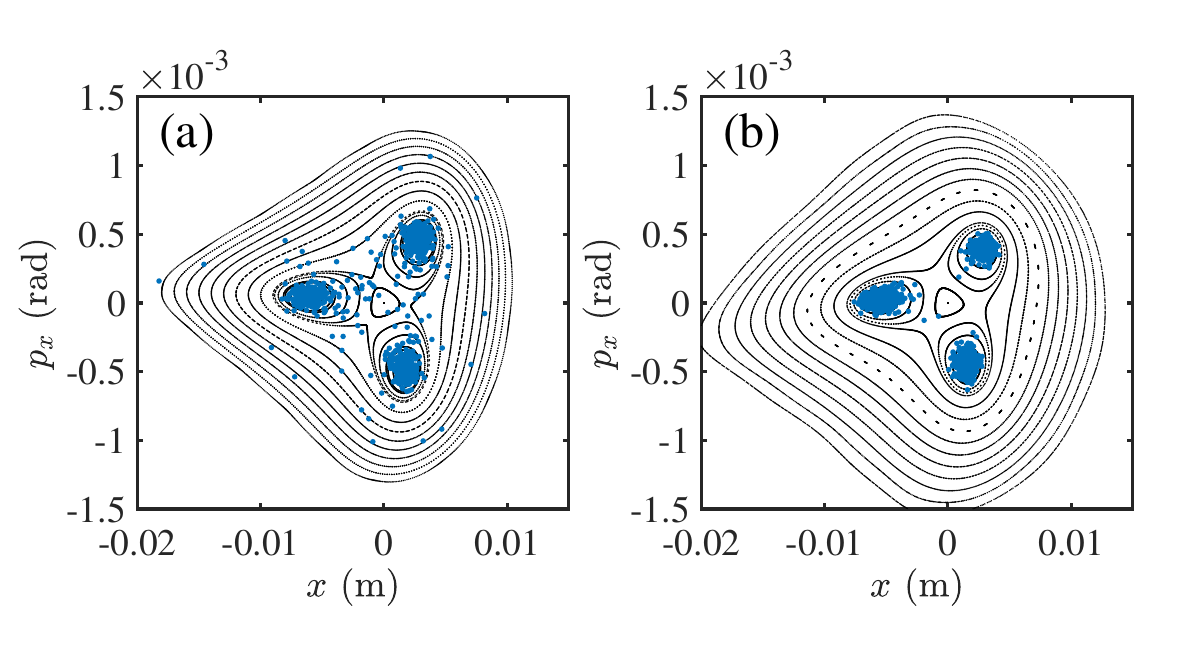}
   \caption{Particles' horizontal phase space before (a) and after (b) the sextupole optimization at $Q_x=0.6651$. The black dots are from tracking 20 particles without radiation damping and excitation. The blue dots show the distribution of 1000 particles after tracking for 40000 turns including radiation effect while applying a sinusoidal kick with $f=259.5$ kHz and $a_p=1$ $\mu$rad.}
   \label{fig:xpx_opt}
\end{figure}

Frequency map analysis \cite{fma:2002, fma:2003} is also conducted for both cases before and after optimization. From Fig.~\ref{fig:fmap} (a) and (c) in the $xy$ plane with small constant energy offset (1$\times$10$^{-5}$), we can see overall diffusion rate ($log_{10}[\sqrt{\Delta\nu_x^2+\Delta\nu_y^2}]$) does not change significantly after optimization as well as the on-energy DA. Note $\Delta\nu_x$ and $\Delta\nu_y$ are the tune difference between the first and second 1000 turns after tracking 2000 turns. However, the diffusion rate near $x=-8$ mm reduces significantly (more blue region), indicating much stabler orbit in the island. This can be seen more clearly in Fig.~\ref{fig:fmap} (d), where the stable island area expands dramatically for off-energy particles after sextupole optimization. And the MA is greatly improved and now limited by CESR RF momentum aperture (6.6$\times 10^{-3}$).

To check whether the sextupole optimization changes the TRIBs behavior, particle tracking simulations are performed on the lattices before and after optimization. As shown in Fig.~\ref{fig:xpx_opt} (a) and (b), the TRIBs contours in the horizontal phase space from tracking 20 on-energy particles without radiation effect are very similar before and after optimization, confirming the DA knobs do not impact TRIBs formation and contours. With radiation damping and excitation included in the tracking, after 40000 turns 1000 particles diffuse into three islands for both cases before and after optimization while applying a sinusoidal kick with the frequency $f=259.5$ kHz and amplitude $a_p=1$ $\mu$rad. The sinusoidal kick is defined as $a_p$sin($2\pi f(n-1)/f_{rev}$), where $a_p$ is the peak amplitude, $f$ is the driving frequency and $n$ is the turn number and $f_{rev}$ is the revolution frequency. However, there are some particles scattered to large amplitudes in phase space before optimization (Fig.~\ref{fig:xpx_opt} (a)) while all particles are well confined in the stable islands after optimization (Fig.~\ref{fig:xpx_opt} (b)). The particles with large amplitudes are presumably lost with the application of the sinusoidal kick, explaining the poor lifetime observed experimentally in the lattice before sextupole optimization. After loading the new sextupoles into CESR, the beam lifetime was indeed improved substantially as the simulations have shown.

\section{Phase space control knobs}\label{phase_con}
As shown in Fig.~\ref{fig:xpx_opt}, the TRIBs phase contours are slightly altered with the sextupole optimization. With other optimization methods that alteration could be more severe when the conditions for TRIBs are not properly constrained. It is thus convenient to construct knobs consisting of a few sextupoles that are capable of fine tuning the TRIBs phase space contours. In addition, these phase-space-control knobs vary the separation of islands such that the x-ray pulses emitted from these particle islands also change, which would be very useful for fine tuning the x-ray positions at the experimental stations. In the following, three kinds of knobs will be discussed to manipulate the TRIBs in phase space.

\subsection{$\phi$ rotation knob}
In Ref.~\cite{chess_tribs:2023}, we discussed a knob which changes $h_{22000}$ at a large scale while keeping $|h_{30000}|$ at minimum change. Five harmonic sextupoles are chosen and grouped to form this sextupole knob. It was discovered the knob can rotate three islands in the horizontal phase space by changing the phase $\phi_0$ of $G$ ($|G|e^{i\phi_0}$), the complex coefficient of the resonance strength at the third-order resonance \cite{chess_tribs:2023}. Besides the angle, the knob also changes the action ($J_{xFP}$) of the fixed points. This knob is created by obtaining the gradients of $|h_{22000}|$ and $|h_{30000}|$ with respect to the five harmonic sextupoles, subtracting the normalized $|h_{30000}|$ components from $|h_{22000}|$ gradients, and normalizing the subtraction results to acquire the knob coefficients. Detailed discussion of this knob including simulation and experimental results can be found in Refs.~\cite{chess_tribs:ipac2023, chess_tribs:2023}.

\begin{figure}
   \centering
   \includegraphics*[width=240pt]{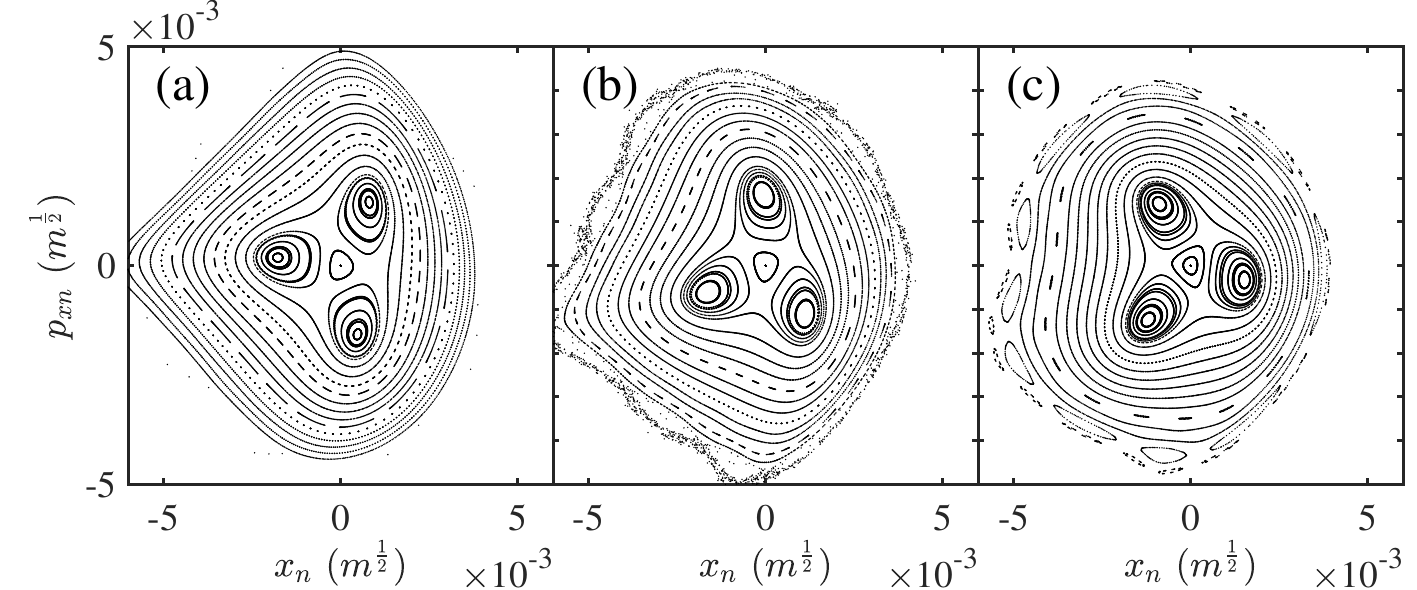}
   \caption{Particles' horizontal phase space from lattices at $Q_x=0.6651$ while setting the $\phi_{\frac{\pi}{6}}$ knob at (a) 0 and (b) 1, and the $\phi_{\frac{\pi}{3}}$ knob at (c) 1. The black dots are the tracking results of 25 particles without radiation damping and excitation.}
   \label{fig:phi_knob}
\end{figure}

In this section we describe another $\phi$ rotation knob that will adjust the angles of the fixed points in phase space but not their actions. Since PTC can extract the detuning coefficients ($\alpha_{xx}$) of ADTS and $G$ precisely so as to predict the fixed points accurately \cite{ef_book:2016,chess_tribs:2023}, PTC codes are integrated into a custom version of Tao optimization program and the parameters $\alpha_{xx}$, $|G|$, $\phi_0$ and $J_{xFP}$ are evaluated as custom datum. If a knob rotates the fixed point in phase space by some amount i.e. $\Delta\phi_x=\frac{\pi}{6}$, the phase of $G$ needs to be changed by $\Delta\phi_0=-3\Delta\phi_x=-\frac{\pi}{2}$ \cite{chess_tribs:2023}. To create this $\phi_{\frac{\pi}{6}}$ knob, $\alpha_{xx}$  remain unchanged and new $G_r$ and $G_i$ targets are chosen such that $|G|$ and $J_{xFP}$ are kept the same but $\phi_0$ changes by $-\frac{\pi}{2}$. $G_r$ and $G_i$ are the real and imaginary parts of $G$. They are used in the optimization instead of $|G|$ and $\phi_0$ because as their use yields faster convergence. All five harmonic sextupoles are included as variables. After optimization, the changes of sextupole strength are calculated as the coefficients of the knob $c_i=\Delta k^{des}_{2i}$, while $i$ is the index of sextupole. When the command value of the $\phi_{\frac{\pi}{6}}$ knob is set to 1, the fixed points in phase space will rotate by the design value $\frac{\pi}{6}$. Similarly, a $\phi_{\frac{\pi}{3}}$ knob can be created.

\begin{figure}
   \centering
   \includegraphics*[width=240pt]{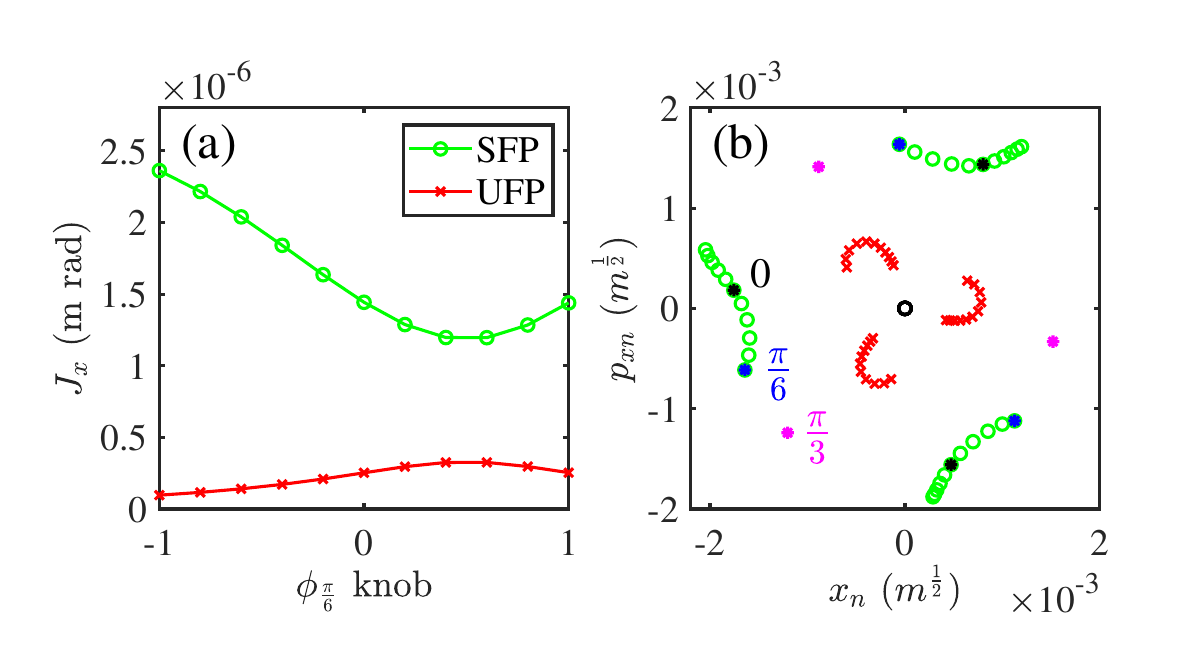}
   \caption{(a) The estimated particle action and (b) the normalized horizontal coordinates of exact SFPs (green circles) and UFPs (red crosses) calculated by PTC codes as a function of the $\phi_{\frac{\pi}{6}}$ knob value while the lattice tune is kept at $Q_x=0.6651$. The black and blue stars are the SFPs while setting $\phi_{\frac{\pi}{6}}$ knob at 0 and 1, respectively. The magenta starts indicate the SFPs while setting $\phi_{\frac{\pi}{3}}$ knob at 1.}
   \label{fig:phi_knob_cal}
\end{figure}

Tracking simulations are performed to verify the knob function. Three cases, no knob change, $\phi_{\frac{\pi}{6}}$ knob value at 1, and $\phi_{\frac{\pi}{3}}$ knob value at 1 are considered in the simulation, while 25 particles are tracked for 1000 turns without radiation damping or excitation included. To view and evaluate the phase angle, we plot the normalized phase space of the tracking results in Fig.~\ref{fig:phi_knob}. The linearly normalized coordinates ($x_n$, $p_{xn}$) are obtained by

\begin{equation}
\begin{pmatrix}
x_n \\
p_{xn} \\
\end{pmatrix} =
\begin{pmatrix}
\frac{1}{\sqrt{\beta_x}} & 0 \\
\frac{\alpha_x}{\sqrt{\beta_x}}  & \sqrt{\beta_x} \\
\end{pmatrix}
\begin{pmatrix}
x \\
p_{x} \\
\end{pmatrix}
 \textrm{,    }
\label{eq:norm}
\end{equation}
where $\alpha_x$ and $\beta_x$ are the Twiss parameters at the first element of lattice.

The phase angle changes of SFPs calculated from Fig.~\ref{fig:phi_knob} for $\phi_{\frac{\pi}{6}}$ at 1 and $\phi_{\frac{\pi}{3}}$ at 1 are 28.6 and 58.0 degrees, respectively, agreeing reasonably well with the design values. When the knob is set to a fraction number $b$, each sextupole of the knob will change by a fraction based on its knob coefficient $\Delta k_{2i}=bc_i=b\Delta k^{des}_{2i}$. This linear change of sextupoles while dialing the knob will not necessarily yield a linear response of the rotation and keep $J_x$ constant. As shown in Fig.~\ref{fig:phi_knob_cal}, we calculated the particle action and the coordinates of the fixed points at different $\phi_{\frac{\pi}{6}}$ knob values. We can see when $\phi_{\frac{\pi}{6}}$ knob changes from 0 to 1, the fixed points rotate fairly linearly from 0 to $\frac{\pi}{6}$ and the particle actions of both SFPs and unstable fixed points (UFPs) change a little. However, when $\phi_{\frac{\pi}{6}}$ knob changes from 0 to $-1$, the angle change of fixed points is less than $\frac{\pi}{6}$ and the actions of the SFPs increase dramatically, indicating the nonlinearity of the $\phi_{\frac{\pi}{6}}$ knob. The $\phi_{\frac{\pi}{3}}$ knob shows more nonlinear response of the angle between 0 to 1 and only the coordinates of SFPs with knob value at 1 are plotted in Fig.~\ref{fig:phi_knob_cal} (b). 

\begin{figure}
   \centering
   \includegraphics*[width=240pt]{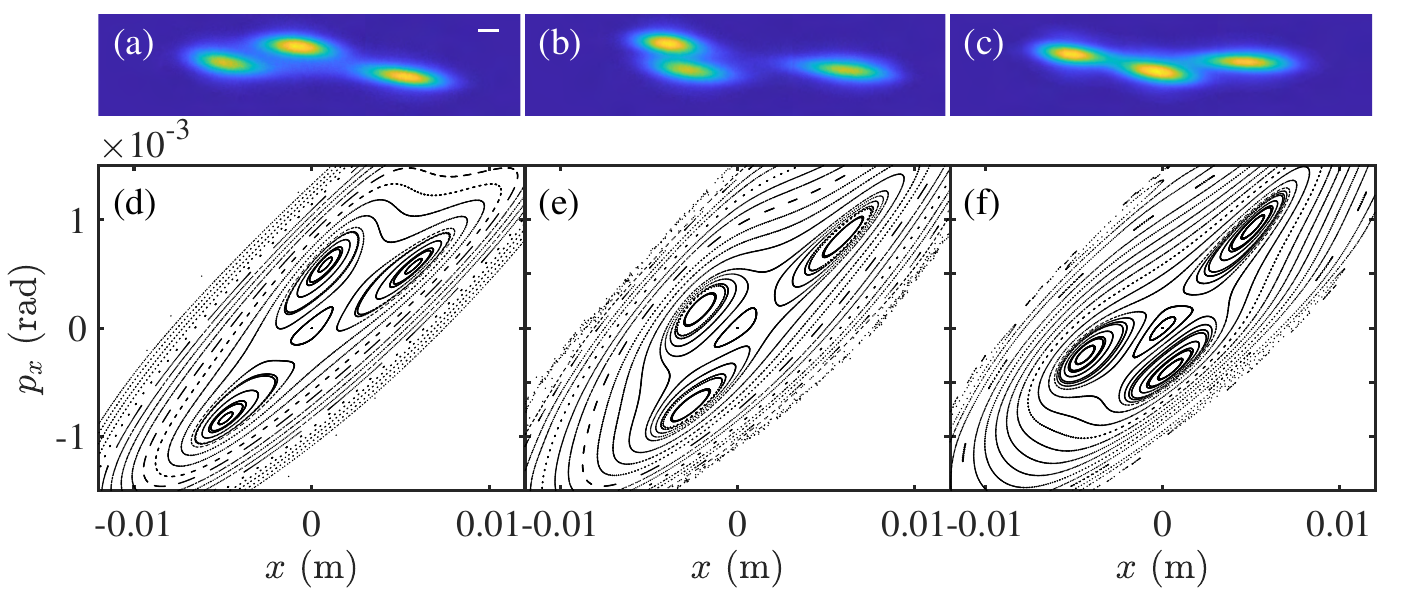}
    \caption{Beam images in $xy$ plane at $Q_x=0.6649$ while setting the $\phi_{\frac{\pi}{6}}$ at (a) 0 and (b) 1, and the $\phi_{\frac{\pi}{3}}$ knob at (c) 1. The bottom plots ((d)-(f)) are particles' horizontal phase space from tracking simulation at the source point of vBSM corresponding to top three cases ((a)-(c)). The white line in (a) indicates a 1-mm scale.}
   \label{fig:phi_knob_exp}
\end{figure}

After $\phi_{\frac{\pi}{6}}$ and $\phi_{\frac{\pi}{3}}$ knobs were loaded in CESR, experiments were conducted to verify the knob properties. For these knobs in CESR, 1000 cu (``computer units") corresponds to the knob value of 1 in simulation. With a single bunch of 0.75 mA stored in CESR,  while the $\phi_{\frac{\pi}{6}}$ knob was varied from 0 to 1000 cu with a step of 200 cu, CCD images of the bunch viewed at vBSM were recorded in a video (See Supplemental Material movie 1). From this video, we can see as the $\phi_{\frac{\pi}{6}}$ knob value increases, the center island moves toward negative $x$ while the left island moves towards positive $x$, eventually they meet and overlap in $x$ when the $\phi_{\frac{\pi}{6}}$ knob was at 1000 cu (Fig.~\ref{fig:phi_knob_exp} (b)). This observation agrees reasonable well with tracking simulations (Fig.~\ref{fig:phi_knob_exp} (d) and (e)). When changing the $\phi_{\frac{\pi}{3}}$ knob, linear rotation of the islands was not observed like the $\phi_{\frac{\pi}{6}}$ knob. However as the $\phi_{\frac{\pi}{3}}$ knob was set at 1000 cu, the center and left island switched their $x$ positions (Fig.~\ref{fig:phi_knob_exp} (c)), consistent with tracking simulation (Fig.~\ref{fig:phi_knob_exp} (f)).

\subsection{$J_x$ knob}
Instead of rotating the islands or fixed points in phase space, knobs that only vary the actions of the fixed points can be created. Changing the actions of both SFPs and UFPs can also vary the island stable area \cite{bessy:2023}. In the design lattice, the particle actions at the SFPs and UFPs ($J_{xSFP}$ and $J_{xUFP}$) are 1.4$\times 10^{-6}$ and 2.2$\times 10^{-7}$ \unit{m\ rad}, respectively, as indicated in Fig.~\ref{fig:phi_knob_cal} (a). Our goal is to bring SFPs and UFPs together at around 7$\times 10^{-7}$ \unit{m\ rad} without rotating them. With the tune fixed at $Q_x=0.6639$, based on the $J_{xSFP}$ and $J_{xUFP}$ target value (7$\times 10^{-7}$ \unit{m\ rad}), we can roughly estimate target values of $\alpha_{xx}$ and $|G|$ as $-$7422 $m^{-1}$ and 0.137 $m^{\frac{1}{2}}$. To keep $\phi_0$ unchanged, both $G_r$ and $G_i$ can be calculated and included in the optimization constraint list. For simplicity, we only include $\alpha_{xx}$ and $|G|$ in the constraints for initial trials. As was done to develop the $\phi$ knob, the customized version of the Tao program is used for optimization. All five harmonic sextuupoles are included as variables in the optimization. Similar procedure as $\phi$ knob is applied to find the $J_x$ knob coefficients after optimization.

\begin{figure}
   \centering
   \includegraphics*[width=240pt]{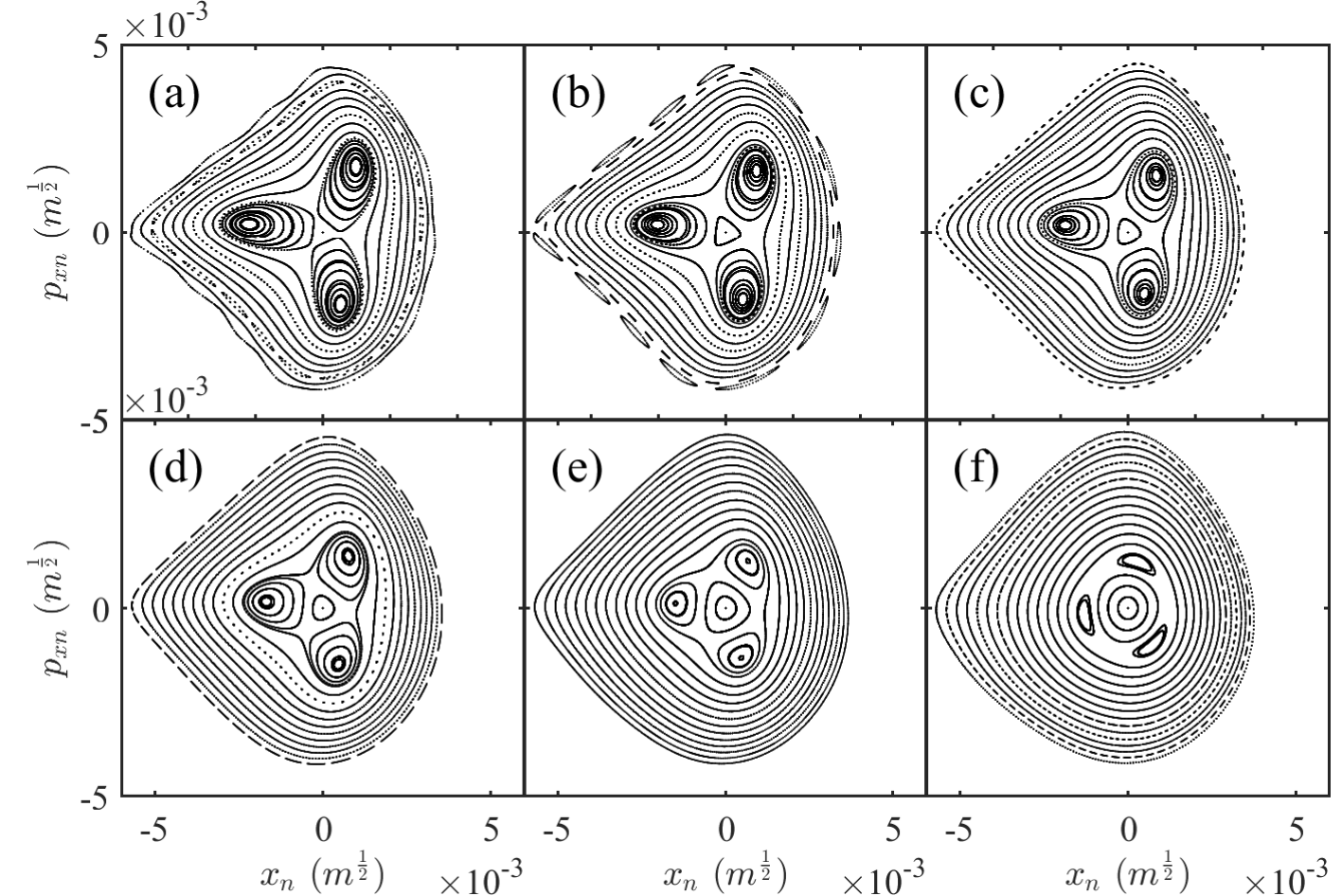}
    \caption{Particles' horizontal phase space from lattices at $Q_x=0.6651$ while setting the $J_x$ knob at (a) $-1.0$, (b )$-0.6$, (c) $-0.2$, (d) 0.2, (e) 0.6, and (f) 1.0. The black dots are the tracking results of 20 particles without radiation damping and excitation.}
   \label{fig:jx_knob}
\end{figure}

\begin{figure}
   \centering
   \includegraphics*[width=240pt]{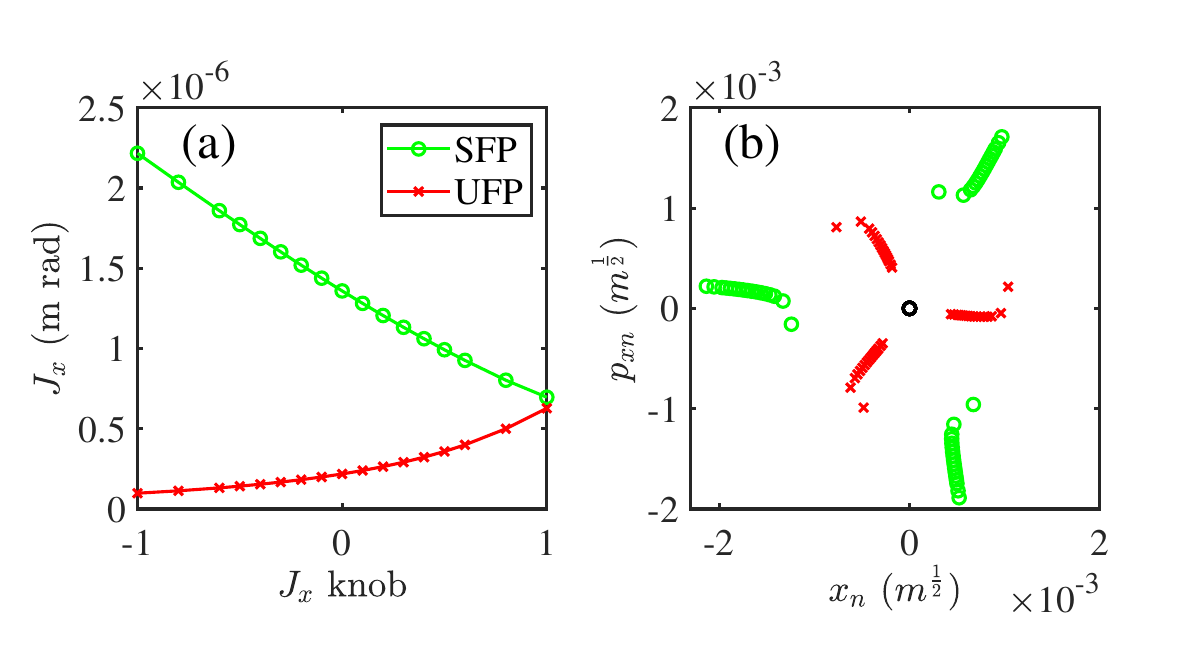}
   \caption{(a) The estimated particle action and (b) the normalized horizontal coordinates of exact SFPs (green circles) and UFPs (red crosses) calculated by PTC codes as a function of the $J_x$ knob value while the lattice tune is kept at $Q_x=0.6651$.}
   \label{fig:jx_knob_cal}
\end{figure}

Similarly, the horizontal phase space contours at different $J_x$ knob settings are obtained from tracking simulation and plotted in Fig.~\ref{fig:jx_knob}, and the particle action and the coordinates of the fixed points with different $J_x$ knob values are calculated and plotted in Fig.~\ref{fig:jx_knob_cal}. When the $J_x$ knob value changes from $-1$ to 1, $J_{xSFP}$ decreases while $J_{xUFP}$ increases as expected (Fig.~\ref{fig:jx_knob_cal} (a)). In addition, the stable island area decreases while the center core area increases when $J_{xSFP}$ and $J_{xUFP}$ change, indicating direct correlation between the island stable area and $J_{xSFP}$ and $J_{xUFP}$. For most knob settings ($-1.0$ to 0.8), both SFPs and UFPs do not rotate much but within $\pm2$ degrees. Only when $J_x$ knob value is 1 (Fig.~\ref{fig:jx_knob} (f)), the SFPs rotate $\sim$13 degrees.

\begin{figure}
   \centering
   \includegraphics*[width=240pt]{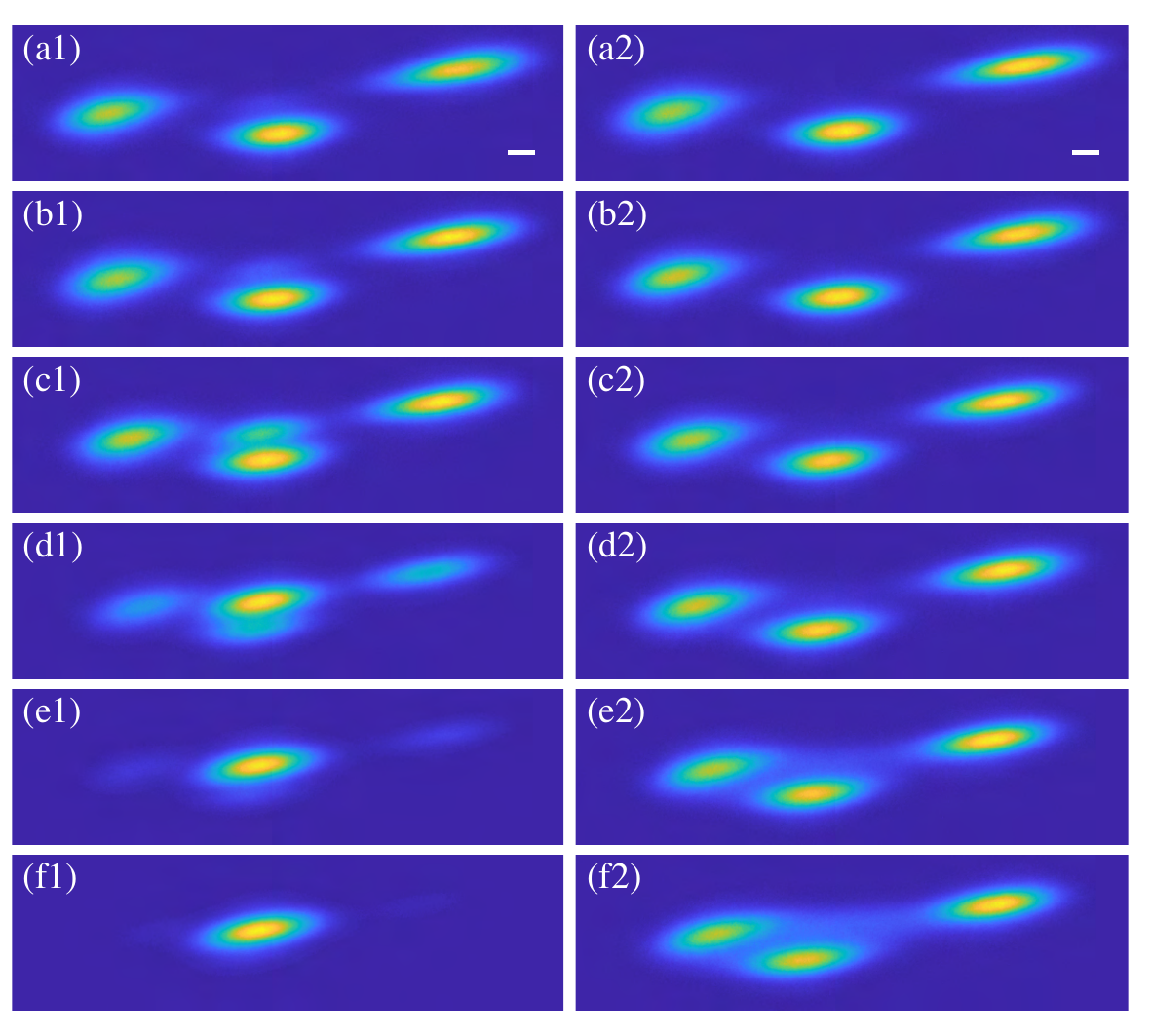}
    \caption{Beam images at $Q_x=0.6649$ without (a1)-(f1) and with (a2)-(f2) applying a sinusoidal kick at $f=259.5$ kHz while the $J_x$ knob was set at (a) $-600$, (b) $-400$, (c) $-200$, (d) 0, (e) 100, and (f) 200\unit{cu}, respectively. The white line indicates a 1-mm scale.}
   \label{fig:jx_knob_exp}
\end{figure}

The $J_x$ knob was loaded in CESR and experiments were then conducted to check the knob behavior. Figure~\ref{fig:jx_knob_exp} shows the recorded vBSM images with islands present at $Q_x=0.6649$ without and with application of a sinusoidal kick while dialing the $J_x$ knob. For the first scenario without kick, while adjusting $J_x$ knob from $-600$ to 200 cu, three islands move closer to the core, indicating the actions of SFPs decrease. Moreover, particles transport more from islands into core as the $J_x$ knob increases. Eventually with $J_x$ knob at 200 cu, all particles stay in the core. This indicates the stable core area greatly expands so that all particles in islands damp down to the core. In addition, three islands positions relative to each other have not changed, showing the angles of SFPs in phase space stay fairly constant, consistent with the simulation results shown in Fig.~\ref{fig:jx_knob_cal} (b). When applying a sinusoidal kick to the beam, particles are all driven into the islands as expected (Fig.~\ref{fig:jx_knob_exp} (a2) to (f2)). 

\subsection{Octupole}\label{oct}
\begin{figure}
   \centering
   \includegraphics*[width=240pt]{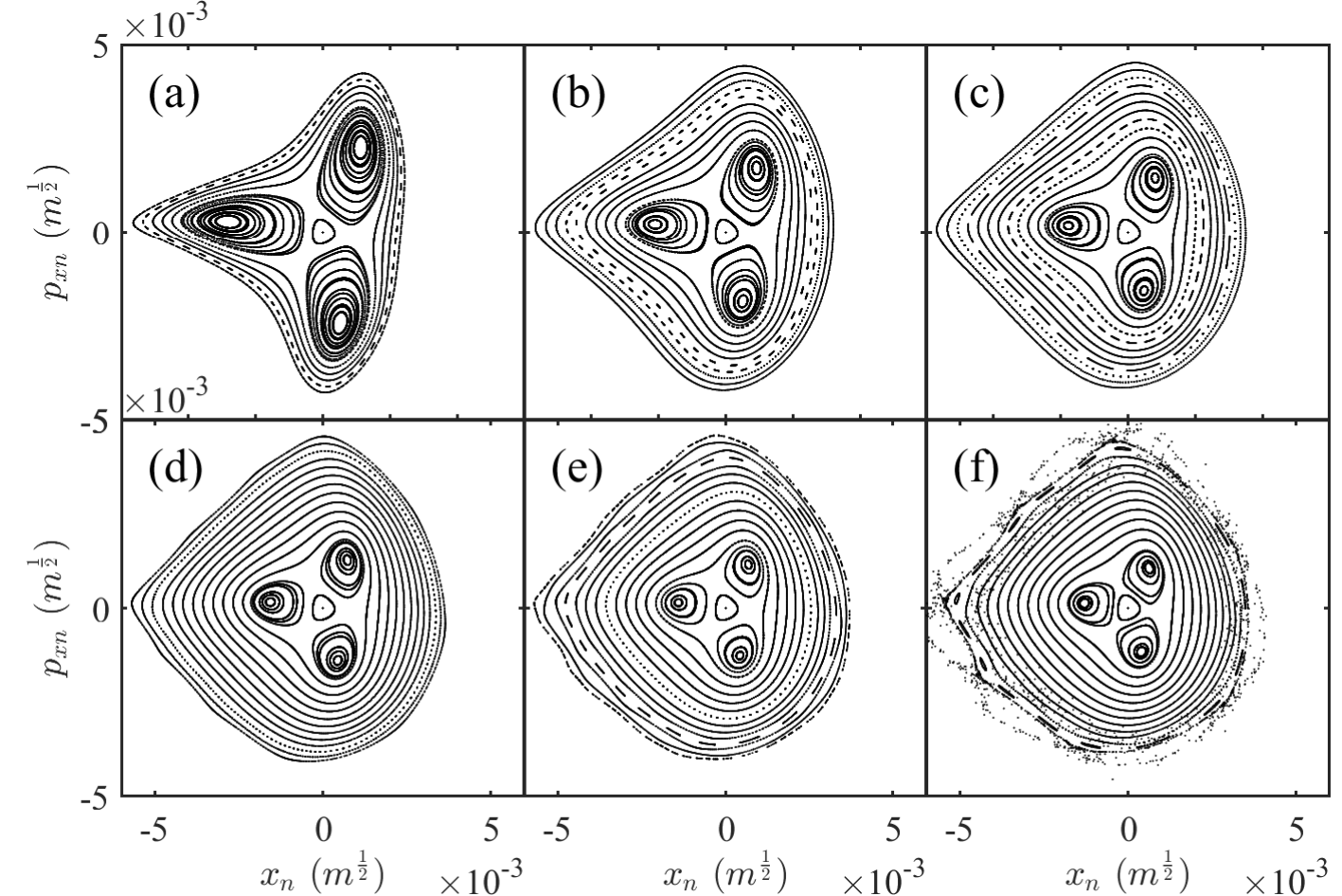}
    \caption{Particles' horizontal phase space from lattices at $Q_x=0.6651$ while setting $k_3$ of two octupoles at (a) $-100$, (b) $-50$, (c) 0, (d) 50, (e) 100, and (f) 150 $m^{-4}$. The black dots are the tracking results of 20 particles without radiation damping and excitation.}
   \label{fig:oct_knob}
\end{figure}

\begin{figure}
   \centering
   \includegraphics*[width=240pt]{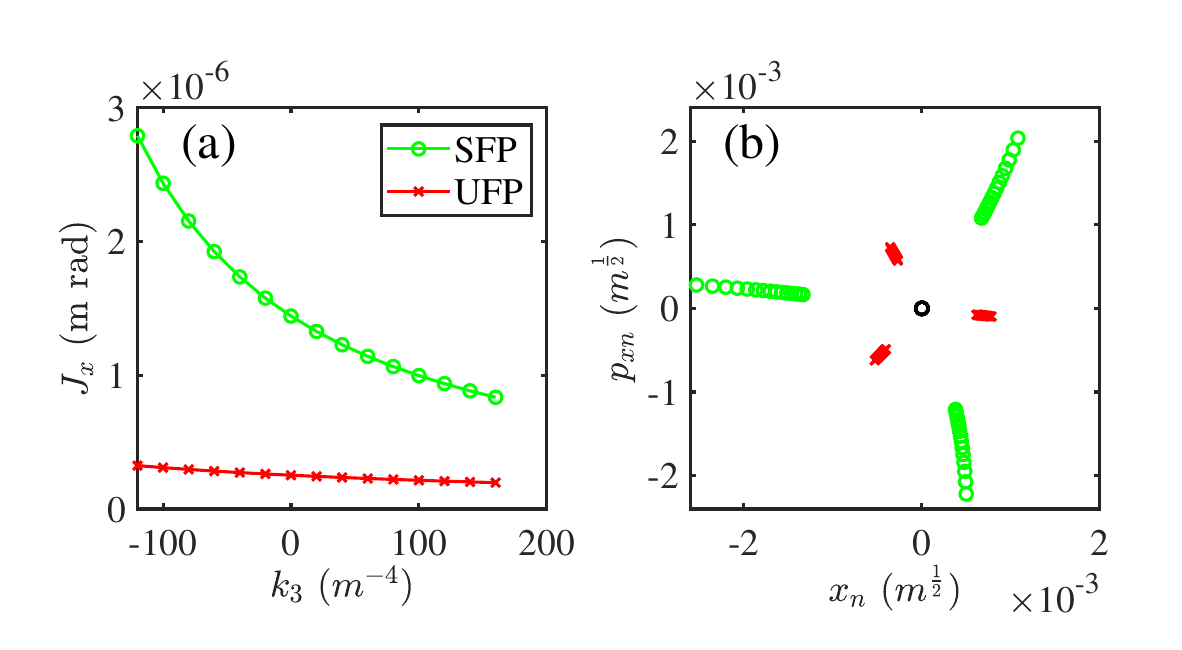}
   \caption{(a) The estimated particle action and (b) the horizontal coordinates of exact SFPs (green circles) and UFPs (red crosses) calculated by PTC codes as a function of the octupole strength ($k_3$) while the lattice tune is kept at $Q_x=0.6651$.}
   \label{fig:oct_knob_cal}
\end{figure}

Octupoles are useful magnets for tuning the coefficients of ADTS ($\alpha_{xx}$) without affecting the third-order resonance strength $G$. The contribution of octupoles to the ADTS is proportional to the strength of octupoles ($k_{3i}$), $\alpha_{xx\_oct}=\frac{1}{16\pi}\sum_{i=1}^{N} \beta^2_{xi}k_{3i}L_i$, where $L_i$ is the length of the $i$-th octupole \cite{oct:pac1991}. In CESR, the two octupoles,  which can be controlled individually, are grouped together for a knob. When the knob is turned, both are varied with the same strength step. Tracking simulations plotted in Fig.~\ref{fig:oct_knob} show when the strength of octupoles increases SFPs get closer to UFPs and the island area decreases. Moreover, the stable core area does not change. This is because UFPs surrounding the core do not change locations in phase space. Figure~\ref{fig:oct_knob_cal} (a) shows the calculated particle actions of SFPs and UFPs as a function of the octupole strength. Indeed, $J_{xSFP}$ decreases while $J_{xUFP}$ remains fairly constant as $k_3$ increases. Since octupoles do not affect $G$, the phase angles of SFPs and UFPs in phase space do not vary when changing the octupoles, which can be seen clearly in Fig.~\ref{fig:oct_knob_cal} (b). 

\begin{figure}
   \centering
   \includegraphics*[width=220pt]{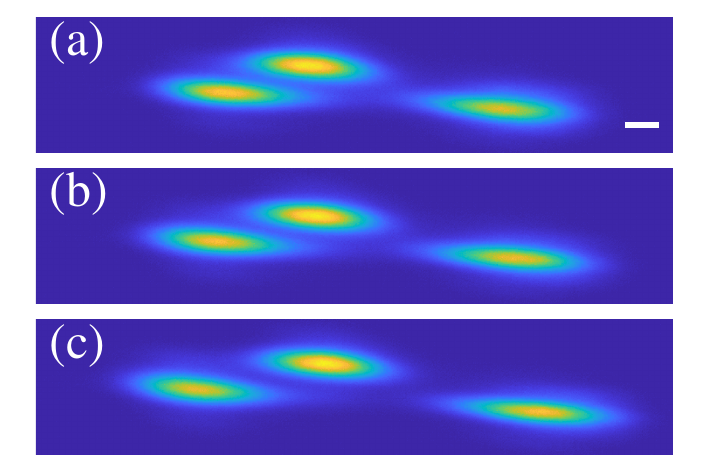}
       \caption{Beam images at $Q_x=0.6651$ with the octupole knob set at (a) $-6000$, (b) 0, and (c) 6000\unit{cu}, respectively. The white line indicates a 1-mm scale.}
   \label{fig:oct_knob_exp}
\end{figure}

Experiments were also conducted to check the octupole effect. As shown in Fig.~\ref{fig:oct_knob_exp}, the islands move further apart and away from the center core without rotation as the octupoles increase from $-6000$ to 6000 cu. The results confirm the calculation in Fig.~\ref{fig:oct_knob_cal} (b) but indicate a wrong sign in current CESR octupole calibration. Further investigations showed the calibration factor was indeed inverted. We will discuss how to use these TRIBs experimental results to calibrate the octupoles in Sec.~\ref{app}.

\section{Particles in a single island}\label{single}
When $Q_x$ is near the third-order resonance but not very close (0.6625), the particles will stay in the core in the absence of a driving kick. When the beam is then driven by a sinusoidal kick with frequency at the core tune $Q_x$, the particles in the core will evenly redistribute into three islands. Similarly, when $Q_x$ approaches closer to the third-order resonance (0.6651),  particles are evenly distributed in three islands due to shrinkage of stable core area. The intensity of x-ray pulse from one evenly-distributed island will be only one third of that from core beam when all particles stay in the core. In addition, there will be x-ray pulses every turn instead of every three turns due to the presence of particles in all three islands. This could be a problem for intensity-thirsty timing experiments. 

To push all particles of a bunch into a single island, one way is to simultaneously apply clearing kicks (two frequencies) for both core and island buckets \cite{bessy:2015}. However, this method involves modifying the harmonic number in the bunch-by-bunch feedback system to apply kick only every third turn or to pause every third turn \cite{bessy:2015}. Intrigued by this kicking scheme, we modified the timing configuration of CESR digital tune tracker \cite{dtt:pac2011} to allow different kick patterns applied to a bunch, such as, applying/pausing 1 kick every third turn. We found particles indeed diffuse into a single island at $Q_x=0.6656$ close to the third-order resonance when applying 1 kick every third turn. With $Q_x=0.6631$ further away from the third-order resonance, pausing 1 kick every third turn drives most particles into a single island but a few particles still remain in the other two islands. This new kicking scheme works but still involves modifying time configuration. 

\begin{figure}
   \centering
   \includegraphics*[width=220pt]{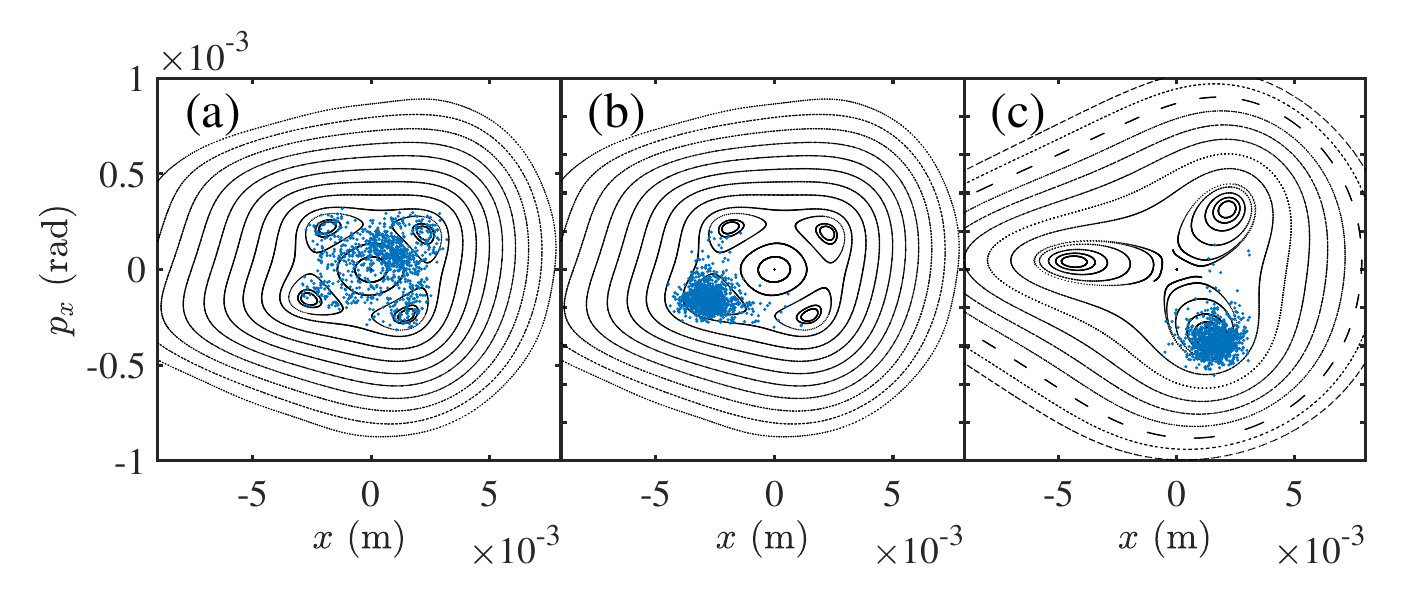}
    \caption{Particles' horizontal phase space at (a) and (b) $Q_x=0.7484$, and (c) $Q_x=0.6661$. The black dots are from tracking 15 particles without radiation damping and excitation. The blue dots show the distribution of 1000 particles after tracking for 40000 turns including radiation effect while applying a sinusoidal kick with a frequency at (a) $f=292.0$ kHz and (b) $f=\frac{3}{4}f_{rev}=292.605$ kHz and amplitude $a_p=1$ $\mu$rad,  (c) $f=\frac{2}{3}f_{rev}=260.093$ kHz and $a_p=2$ $\mu$rad, respectively.}
   \label{fig:single_sim}
\end{figure}

While studying the TRIBs near 4$\nu_x$, we found another mechanism to drive all particles of a bunch into a single island without modifying timing configurations of either tune tracker or feedback system \cite{chess_tribs:ipac2024}. Tracking simulation with RF, radiation damping and excitation turned on is implemented to check how particles form islands near 4$\nu_x$ while applying a sinusoidal kick. Starting with an initial distribution with the design emittance $\epsilon_x=30$ nm rad, 1000 particles are tracked for 40000 turns ($\sim$8 damping times). With the driving frequency at the core tune $Q_x=0.7484$ ($f=292.0$ kHz) and $a_p=0.5$ $\mu$rad, particles are pushed from the core to occupy all four islands (Fig.~\ref{fig:single_sim} (a)). Interestingly, when the driving frequency is set at the exact $4\nu_x$ tune $Q_x=\frac{3}{4}$ ($f=\frac{3}{4}f_{rev}=292.605$ kHz), all particles are driven into a single islands (Fig.~\ref{fig:single_sim} (b)). Similarly for the third-order resonance at $Q_x=0.6661$, while applying a sinusoidal kick with a frequency at the exact $3\nu_x$ tune ($f=\frac{2}{3}f_{rev}=260.093$ kHz), particles are driven into a single island as well (Fig.~\ref{fig:single_sim} (c)).

\begin{figure}
\centering
\includegraphics*[width=220pt]{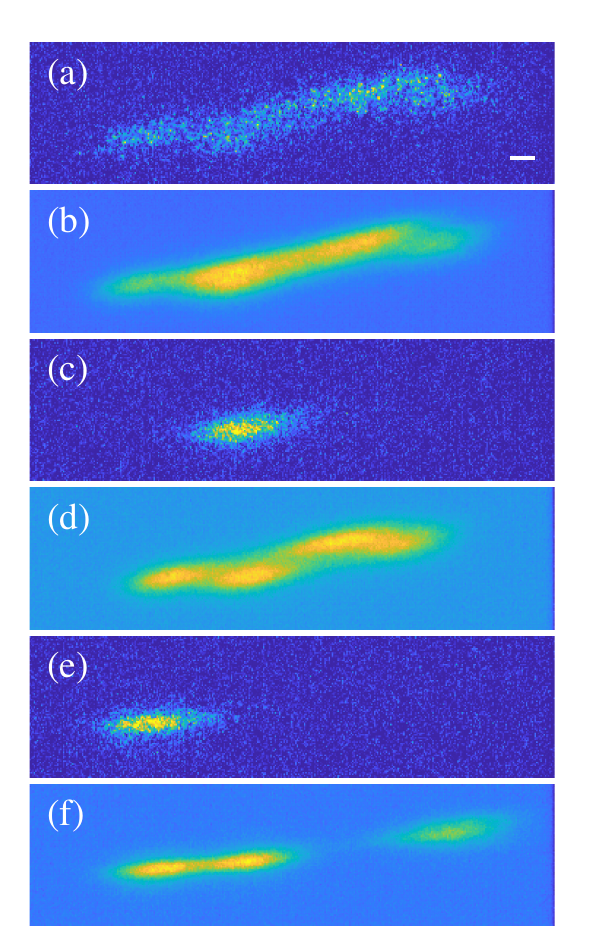}
\caption{(a) Single-shot frame and (b) averaged image over 100 frames taken at $Q_x=0.7474$ ($291.586$ kHz) with a sinusoidal kick at $f=291.87$ kHz. (c) Single-shot frame and (d) averaged image over 100 frames taken at $Q_x=0.7487$ ($292.09$ kHz) with a sinusoidal kick at $f=292.6$ kHz. (e) Single-shot frame and (f) averaged image over 100 frames taken at $Q_x=0.6661$ ($259.88$ kHz) with a sinusoidal kick at $f=260.09$ kHz. The white line indicates a 1-mm scale.}
\label{fig:single_exp}
\end{figure}

Since the images obtained by the CCD camera are integrated over many turns, they could not tell whether the particles occupy one single island or all four islands. Thus a fast-gated camera is used to acquire the single-shot (one-turn) bunch profile \cite{vbsm:2017}. Figure~\ref{fig:single_exp} (a) and (b) show a single-shot frame and the averaged image of 200 frames at $Q_x=0.7474$ (291.586 kHz) with a sinusoidal kick at $291.87$ kHz. Consistent with the simulation (Fig. \ref{fig:single_sim} (a)), particles occupy all four islands and the core. However at the same tune but driving with a different frequency at 292.6 kHz, the single-shot image in Fig.~\ref{fig:single_exp} (c) indicates all particles are in one single island, confirming the simulation result in Fig.~\ref{fig:single_sim} (b), while the averaged image (Fig.~\ref{fig:single_exp} (d)) still shows four islands. Figure \ref{fig:single_exp} (e) and (f) show a single-shot frame and the averaged image of 200 frames near $3\nu_x$ at $Q_x=0.6662$ (259.88 kHz) with the driving frequency at $260.09$ kHz, indicating all particles are in one single island and consistent with the simulation result in Fig. \ref{fig:single_sim} (c). Three videos showing different particle distributions under different frequencies of sinusoidal kick can be found in the Supplemental Material movie 2 to 4. 

\begin{figure}
\centering
\includegraphics*[width=200pt]{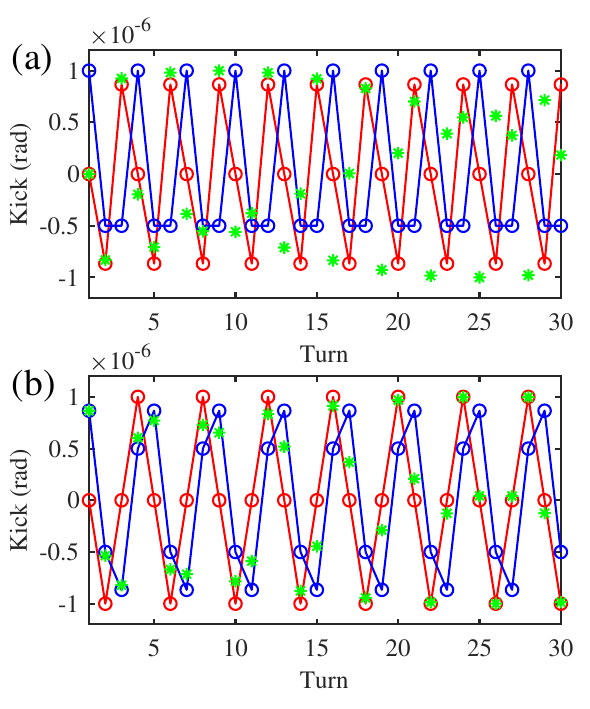}
 \caption{The sinusoidal kicks as a function of turn. In (a), the red and blue circles are the kicks with the frequency at the resonant tune $Q_x=\frac{2}{3}$ (260.093 kHz) with a phase of $0$ and $\frac{\pi}{2}$, respectively. The green stars show the kicks with $f=256$ kHz. In (b), the red and blue circles are the kicks with the frequency at $Q_x=\frac{3}{4}$ (292.605 kHz) with a phase of $0$ and $\frac{\pi}{3}$, respectively. The green stars show the kicks with $f=290$ kHz.}
\label{fig:single_kick}
\end{figure}

To understand why the sinusoidal kick at the exact 3$\nu_x$ or 4$\nu_x$ frequencies can drive particles into one single island, we plot the sinusoidal kick as a function of turn at different frequencies in Fig.~\ref{fig:single_kick}. The sinusoidal kick amplitude is defined as 
\begin{equation}
A_s = a_p \sin(\frac{2\pi f (n-1)}{f_{rev}} + \phi_s)
\ \label{eq:sin_kick}  \textrm{,}
\end{equation}
where $\phi_s$ is the phase. In Fig.~\ref{fig:single_kick}, $a_p=1$ $\mu$rad is used for all frequencies. When the kick frequency is set at $\frac{2}{3}f_{rev}$ or $\frac{3}{4}f_{rev}$ , Eq.~\ref{eq:sin_kick} indicates the period of the sinusoidal kick is 3 or 4 turns, respectively, as Fig.~\ref{fig:single_kick} (a) and (b) show. In other words, the beam receives the same kick every 3 or 4 turns while the kick amplitude each turn depends on $\phi_s$. For example, for $f=\frac{2}{3}f_{rev}$ (Fig.~\ref{fig:single_kick} (a)),  there is no kick applied when $\phi_s=0$ at turn $n=1,4,7...$ but full kick 1 $\mu$rad when $\phi_s=\frac{\pi}{2}$. When $f$ is not at exact 3$\nu_x$ or 4$\nu_x$ or any other resonant frequencies, the kick varies every turn (see the green stars in Fig.~\ref{fig:single_kick}). Due to different topology in phase space, the particles in different islands will have different behavior after the kick. In other words, the particles in one island may be easily kicked out of its stable area so as to diffuse into other region while the particles in other two islands are more resistant (stable) to the kick. Because the kicks are constant every three turns, after damping for many turns, the particles eventually diffuse into one region in phase space such that they are stable and less affected by the three kicks every three turns. If the stable region happens to coincide with one island, the particles will appear in one single island as we observed. If the stable region coincides with two islands, we will see particles appear in two islands.  

\begin{figure}
\centering
\includegraphics*[width=240pt]{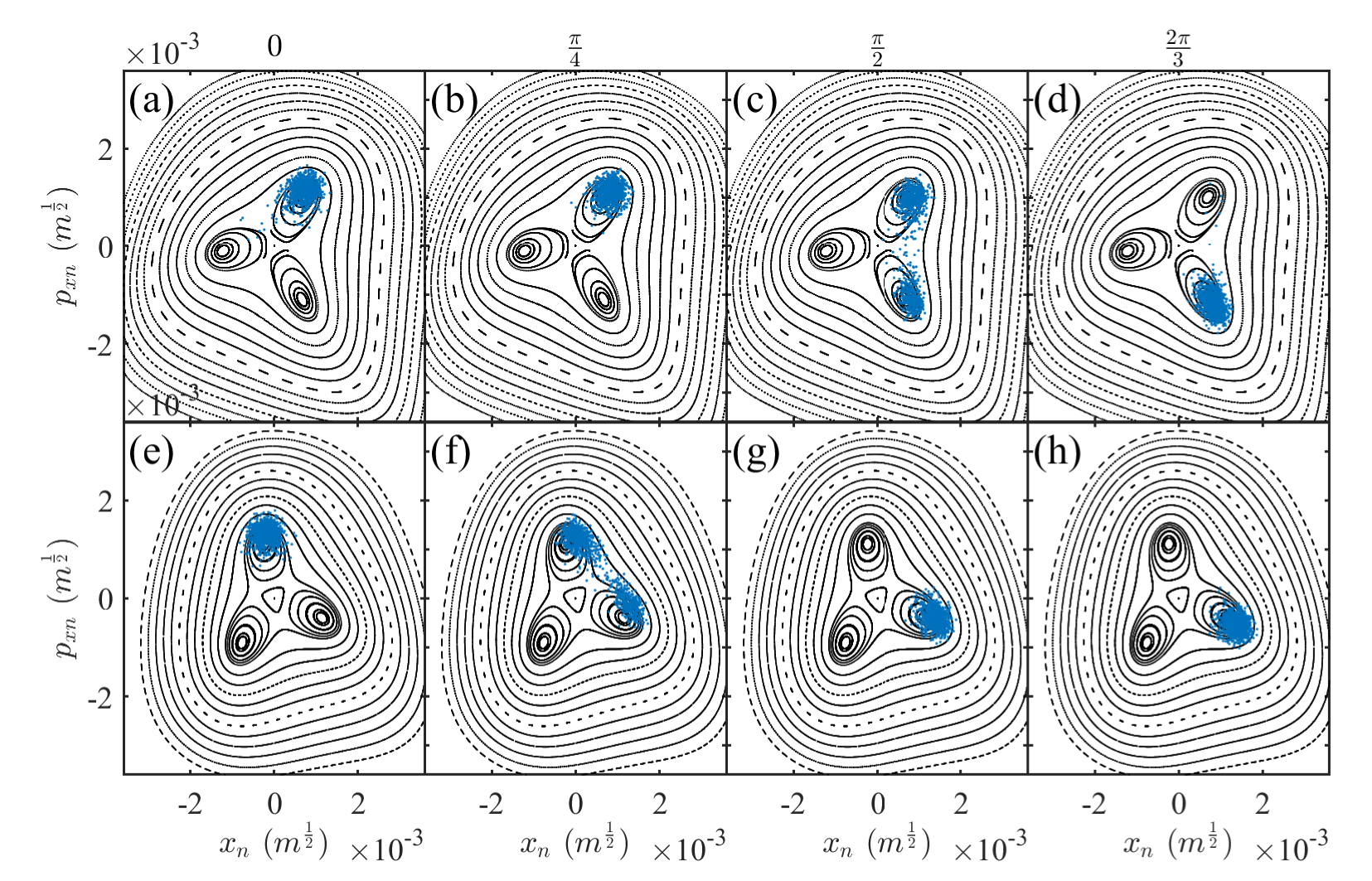}
 \caption{Particles' horizontal phase space with the lattice tune of $Q_x=0.6661$ (259.88 kHz) and with ADTS knob value set at $0$ ((a)-(d)) and $-1$ ((e)-(h)), respectively. The blue dots show the distribution of 1000 particles after tracking for 40000 turns including radiation effect while applying a sinusoidal kick with $f=\frac{2}{3}f_{rev}=260.093$ kHz and $a_p=2$ $\mu$rad and $\phi_s$ at $0$, $\frac{\pi}{4}$, $\frac{\pi}{2}$, and $\frac{2\pi}{3}$, respectively.}
\label{fig:3nux_sim}
\end{figure}

If the above explanation about driving particles into single island with resonant sinusoidal kick is correct, the observation of one island or two islands will depend not only on the topology of phase space but also on $\phi_s$, which determines the three kicks in one kick period. In the simulations discussed above, we have always set $\phi_s=0$ (Fig.~\ref{fig:single_sim}). Here we perform tracking simulations on two lattices with ADTS knob value set at $0$ and $-1$ while applying the sinusoidal kick with $f=\frac{2}{3}f_{rev}$ and $a_p=2$ $\mu$rad but with different phases ($0$, $\frac{\pi}{4}$, $\frac{\pi}{2}$, and $\frac{2\pi}{3}$). The results are shown in Fig.~\ref{fig:3nux_sim}. For the lattice with zero ADTS knob value, after 40000 turns all 1000 particles are driven into a single island with $\phi_s=0$ (Fig.~\ref{fig:3nux_sim} (a)) but into two islands with $\phi_s=\frac{\pi}{2}$ (Fig.~\ref{fig:3nux_sim} (c)). For the lattice with ADTS knob value at $-1$, three islands in phase space rotate $\sim$46.4 degrees \cite{chess_tribs:2023}. The tracking simulation with this lattice shows 1000 particles are driven into two islands with $\phi_s=\frac{\pi}{4}$ (Fig.~\ref{fig:3nux_sim} (f)). The change of $\phi_s$ for two islands is $-45$ degree, consistent with the rotation of islands. 

\begin{figure}
   \centering
   \includegraphics*[width=220pt]{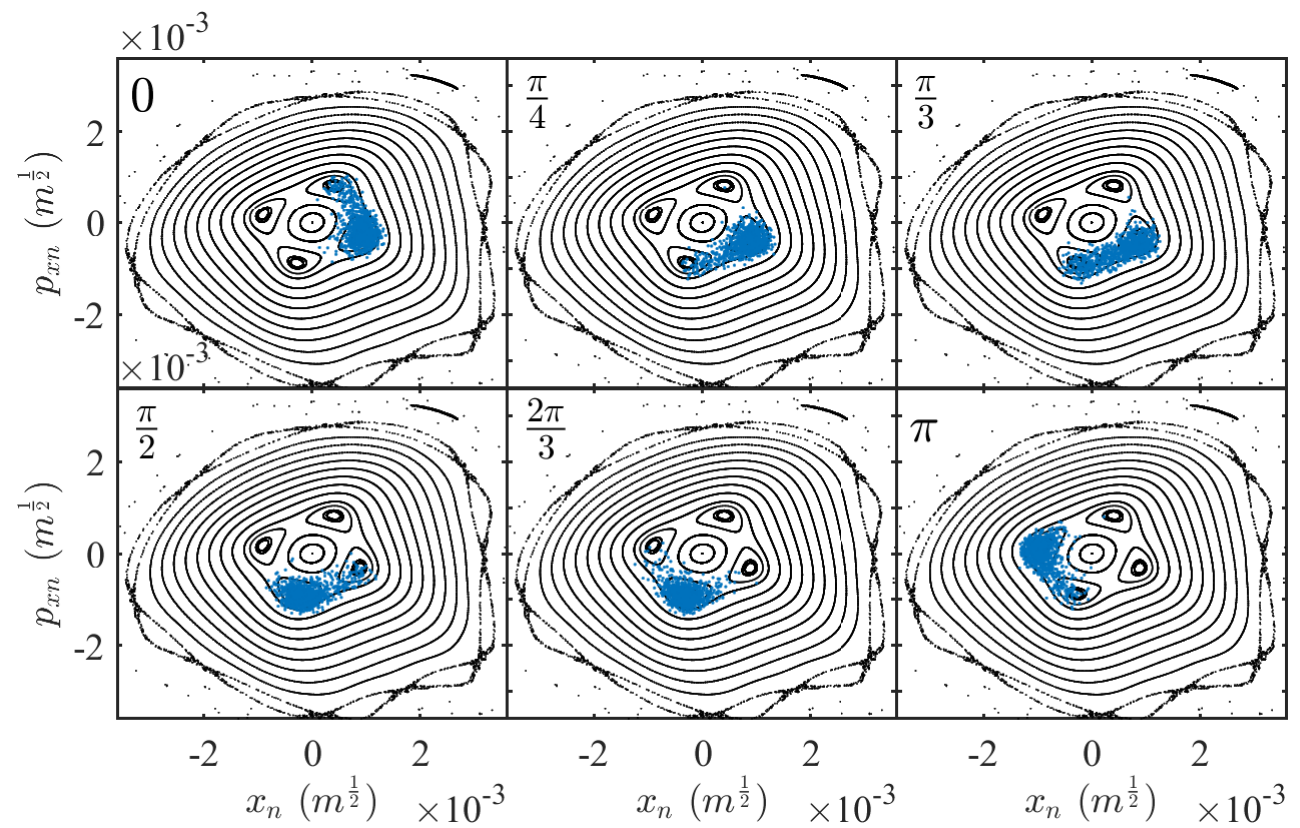}
   \caption{Particles' horizontal phase space with the lattice tune of $Q_x=0.7484$ (292.0 kHz). The blue dots show the distribution of 1000 particles after tracking for 40000 turns including radiation effect while applying a sinusoidal kick with $f=\frac{3}{4}f_{rev}=292.605$ kHz and $a_p=1$ $\mu$rad and $\phi_s$ at $0$, $\frac{\pi}{4}$, $\frac{\pi}{3}$, $\frac{\pi}{2}$, $\frac{2\pi}{3}$, and $\pi$, respectively}
   \label{fig:4nux_sim}
\end{figure}

Tracking simulations are also performed with the lattice tune near 4$\nu_x$ at $Q_x=0.7484$ (292.0 kHz) with $f=\frac{3}{4}f_{rev}$ and $a_p=1$ $\mu$rad and with different $\phi_s$ as shown in Fig.~\ref{fig:4nux_sim}. Similarly to 3$\nu_x$, varying $\phi_x$ shifts particles occupying one island to two. The phase period changes from $\frac{2\pi}{3}$ for 3$\nu_x$ to $\frac{\pi}{2}$ for 4$\nu_x$. These simulation results confirm our speculation about the driving mechanism of particles in one island. Currently we do not have the control of the kick phase. When turning on the sinusoidal kick, the phase could be randomly picked. During experiments, we found particles are driven into a single island for most cases. However, occasionally we did observe that some particles still remained in another island, indicating a different phase. In the future, it would be interesting and worth doing to have the control of $\phi_s$ to further validate the simulation results experimentally.

\section{Application and discussion}\label{app}
In the TRIBs mode, besides the central closed orbit (core), there exists a second closed orbit (island), in which the beam travels 3 or 4 turns before returning to its original position in phase space depending on the horizontal tune near a 3rd-order or 4th-order resonance line. Thus, the frequency of x-ray pulses from the particle beam populating an island in the second orbit decreases by 3 or 4 times which expands the opportunities for timing experiments \cite{bessy:2019}. In addition, x-rays from beams in two orbits are also separated in both horizontal and vertical planes. This unique way of separating x-ray pulses in both time and space attracts more attention from x-ray users for timing experiments. Recently scientists demonstrated a unique method of flipping the helicity of x-rays at very high rates (2~\unit{ns}) by passing two horizontally-separated bunches through a twin elliptical undulator in TRIBs mode \cite{hel_switch:2020}. Another proof-of-principle experiment was demonstrated using near-edge x-ray spectroscopy. The combination of transverse offset of the source points from TRIBs and direct imaging monochromators allowed tunable energy shifts of a few eV \cite{sci_rep:2022}. Different source spots defined by the TRIBs three turn closed orbit enable one to normalize resonant absorption signals on the MHz scale. 

On the accelerator side, people have utilized the resonant nature of nonlinear resonance for particle extraction for many years. Resonant extraction is a well established technique for gradual removal of circulating beams from storage rings and synchrotrons, and indeed was standard operating procedure for the Cornell Synchrotron when it was used for fixed target high energy physics experiments \cite{cornell_sync:pac1971}. Other examples are described in Sec.~\ref{intro}. Here we discuss two other applications in CESR that have not been realized elsewhere: (a) calibration of octupole elements and (b) gain calibration of beam position monitors (BPM). Recall that to measure the octupole effect (Sec.~\ref{oct}), the CCD images with the vBSM were acquired as a function of octupole strength. As shown in Fig.~\ref{fig:oct_knob_exp}, the islands move apart while increasing the octupole knob value from $-6000$ to 6000 cu. The simulation in Fig.~\ref{fig:oct_knob} shows the islands move closer while increasing the octupole strength. These results indicate that there is a sign error in the calibration factor. Indeed by comparing experimental measurements and simulation results, we can extract the calibration factors directly. First, the images are analyzed to find the horizontal locations of the core and islands at three different octupole settings ($-6000$, 0, $6000$ cu) and then the changes of the island horizontal positions respect to the octupole change ($\partial(mm)/\partial(cu)$) are calculated. Second, three SFPs at the vBSM source points are calculated at three different octupole strength ($-100$, 0, $100$ $m^{-4}$) and the change of island positions respect to octupole strength are found ($\partial(mm)/\partial(m^{-4})$). Then finally, the calibration factor is obtained as $-6.1\times10^{-3}$ $m^{-4}/cu$ at 6 GeV, that is 1000 cu of octupole knob equals to $k_3=-6.1$ $m^{-4}$, consistent with the design value $4.4\times10^{-3}$ $m^{-4}/cu$ at 6 GeV \cite{oct:cbn81}.

Recently, a beam-based method was developed to calibrate the relative gains of BPM pick-up electrodes at CESR \cite{gain:ipac2024}. This method solves a system of equations for different beam positions and simultaneously for the relative gains by comparison with the nonlinear response map of the pick-up electrodes as a function of the beam position \cite{bpm_map:2005}. At a given BPM, multiple beam positions ($>4$) with horizontal and vertical spatial separation greater than 500 $\mu$m are required in order to construct the gains with suitable precision. Currently, the beam is displaced locally using steering magnets to acquire 9 positions at a given BPM and it takes about 15 minutes to collect data for a single or few BPMs, making it impractical to quickly calibrate all the 100 BPMs. In TRIBs lattice near a 3$\nu_x$ or 4$\nu_x$ resonance, a bunch can be driven into a single island as discussed above and circulates between the islands in 3 or 4 turns so that 3 or 4 unique beam positions at all BPMs can be acquired at once when collecting turn-by-turn (TBT) data. By changing the tune $Q_x$, $xy$ coupling, or any phase-control knob so as to adjust the island locations in phase space, 3 or 4 additional positions at all BPMs can be acquired once again. For a proof-of-principle experiment, from changing the $xy$ coupling with a skew-quadrupole, we have obtained 7 beam positions (core and islands) at all BPMs. Preliminary analysis of these TBT data using the aforementioned method shows the relative gains of some BPMs agree reasonably well with those acquired from analyzing the position data from local bumping while some BPMs have less agreement. There are several possible explanations for the disagreement such as limited vertical separations between beam positions ($<$500 $\mu$m), noisy tune modulation, and existence of residual particles in the other island, etc. Further detailed data analysis are in progress.

\begin{figure}
   \centering
   \includegraphics*[width=210pt]{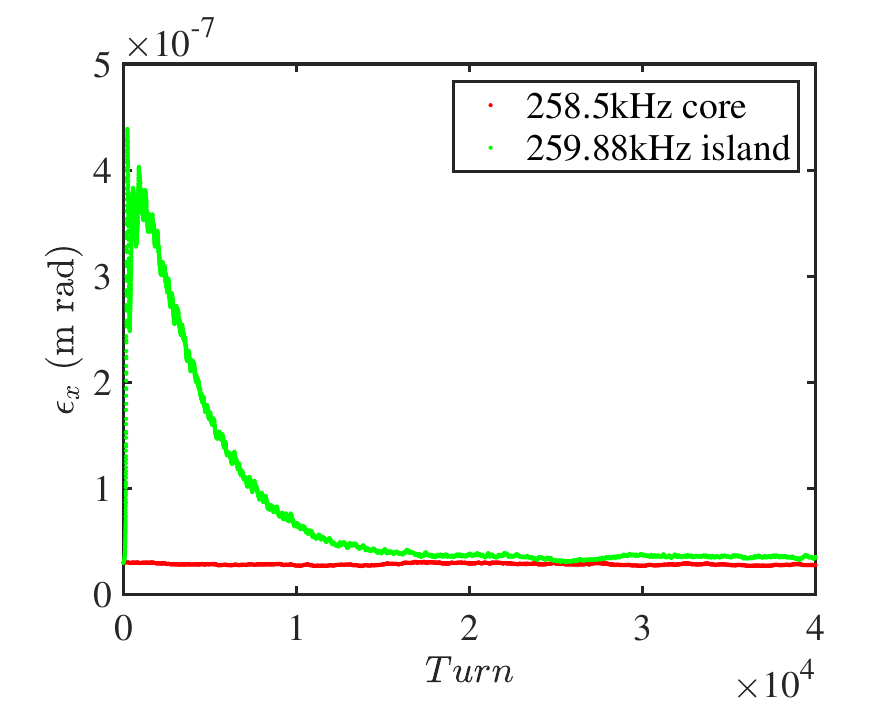}
   \caption{The calculated horizontal emittance ($\epsilon_x$) as a function of turn from tracking 1000 particles with radiation damping and excitation effect. The red curve shows the result with the lattice tune at $Q_x=0.6626$ (258.5 kHz) without applying kicks. The green curve shows the result with $Q_x=0.6661$ (259.88 kHz) while applying a sinusoidal kick with $f=\frac{2}{3}f_{rev}=260.093$ kHz) and $a_p=2$ $\mu$rad and $\phi_s=\frac{\pi}{4}$.}
   \label{fig:emit}
\end{figure}

As discussed in Sec.~\ref{single}, to drive all particles of a bunch into one single island, a sinusoidal horizontal kick is necessarily applied to the bunch. This continuous kick could increase the emittance of the bunch in the island. Tracking simulations are performed to understand the effect. The emittances of the bunch consisting of 1000 particles are calculated every turn based on its sigma matrix while tracking for 40000 turns with radiation damping and excitation included \cite{wolski:2006}. Figure~\ref{fig:emit} shows the horizontal emittance of a bunch in the core at $Q_x=0.6626$ (258.5 kHz) without applying kicks and a bunch at $Q_x=0.6661$ (259.88 kHz) while applying a sinusoidal kick with the frequency at $f=\frac{2}{3}f_{rev}$ (260.093 kHz) and $\phi_s=\frac{\pi}{4}$. Without the sinusoidal kick, the particles stay in the core and the emittance constant, turn to turn, at $\sim$28.6 nm rad. With applying the sinusoidal kick, the particles in the core diffuse into islands, reflected in the increase of the calculated emittance. Once particles start to converge into one island, $\epsilon_x$ gradually decreases. After 20000 turns when all particles are driven into the single island, the emittance of the bunch converges to $\sim$35.5 nm rad, which is about 24$\%$ more than that of the bunch in core. This non-negligible increase of $\epsilon_x$ will reduce the x-ray flux from the bunch by 24$\%$ as well, which may or may not be a problem depending on the specific x-ray timing experiment.

In conclusion, we have discussed a new method to improve the DA and MA of the TRIBs lattice while preserving the TRIBs properties. Different knobs, $\phi$ and $J_x$ knobs and octupoles, are introduced to manipulate the phase space so as to adjust the islands or fixed point locations. These could be very useful for the fine-tuning of x-ray pulses. A new scheme to drive all particles into one single island is also described. Several new applications of TRIBs are discussed.

\begin{acknowledgments}
The authors thank David Rubin for valuable discussion and comments on this paper, Robert Meller for useful discussion and assistance with the tune tracker, David Sagan for his assistance with BMAD, P. Nishikawa for his help of PTC codes, Joel Brock and Ernest Fontes for supporting this project, and CESR operation group for their support during machine study shifts. This research was supported by NSF award DMR-1829070.
\end{acknowledgments}

\bibliography{tribs}

\end{document}